\documentclass[sw]{iosart2x}

\pubyear{0000}
\volume{0}
\firstpage{1}
\lastpage{1}

\begin{document}

\begin{frontmatter}

%\pretitle{}
\title{MAMBO: a lightweight ontology for multiscale materials and applications}
\runtitle{MAMBO: a lightweight ontology for multiscale materials and applications}
%\subtitle{}

% For one author:
%\author{\inits{N.}\fnms{Name1} \snm{Surname1}\ead[label=e1]{first@somewhere.com}}
%\address{Department first, \orgname{University or Company name},
%Abbreviate US states, \cny{Country}\printead[presep={\\}]{e1}}

% Two or more authors:
\begin{aug}
\author[A,B]{\inits{F.}\fnms{Fabio} \snm{Le Piane}\ead[label=e1]{fabio.lepiane@ismn.cnr.it}}
\author[A]{\inits{M.}\fnms{Matteo} \snm{Baldoni}\ead[label=e2]{matteo.baldoni@ismn.cnr.it}}
\author[B]{\inits{M.}\fnms{Mauro} \snm{Gaspari}\ead[label=e3]{mauro.gaspari@unibo.it}}
\author[A]{\inits{F.}\fnms{Francesco} \snm{Mercuri}\ead[label=e4]{francesco.mercuri@cnr.it}%
\thanks{Corresponding author. \printead{e4}.}}
\address[A]{DAIMON Lab, Istituto per lo Studio dei Materiali Nanostrutturati (ISMN), \orgname{Consiglio Nazionale delle Ricerche (CNR)}, \cny{Italy}\printead[presep={\\}]{e1,e2,e4}}
\address[B]{Department of Computer Science, \orgname{University of Bologna}, \cny{Italy}\printead[presep={\\}]{e3}}
\end{aug}

%\begin{review}{editor}
%\reviewer{\fnms{First} \snm{Editor}\address{\orgname{University or Company name}, \cny{Country}}}
%\reviewer{\fnms{Second} \snm{Editor}\address{\orgname{First University or Company name}, \cny{Country}
%    and \orgname{Second University or Company name}, \cny{Country}}}
%\end{review}
%\begin{review}{solicited}
%\reviewer{\fnms{First} \snm{Solicited reviewer}\address{\orgname{University or Company name}, \cny{Country}}}
%\reviewer{\snm{anonymous reviewer}}
%\end{review}
%\begin{review}{open}
%\reviewer{\fnms{First} \snm{Open Reviewer}\address{\orgname{University or Company name}, \cny{Country}}}
%\end{review}

\begin{abstract}
Advancements of both computational and experimental tools have recently led to significant progress in the development of new advanced and functional materials, paralleled by a quick growth of the overall amount of data and information on materials.
However, an effective unfolding of the potential of advanced and data-intensive methodologies requires systematic and efficient methods for the organization of knowledge in the context of materials research and development.
Semantic technologies can support the structured and formal organization of knowledge, providing a platform for the integration and interoperability of data.
In this work, we introduce the Materials and Molecules Basic Ontology (MAMBO), which aims at organizing knowledge in the field of computational and experimental workflows on molecular materials and related systems (nanomaterials, supramolecular systems, molecular aggregates, etc.).
Linking recent efforts on ontologies for materials sciences in neighboring domains, MAMBO aims at filling gaps in current state-of-the-art knowledge modelling approaches for materials development and design targeting the intersection between the molecular scale and higher scale domains.
With a focus on operational processes, lightweight, and modularity, MAMBO
enables extensions to broader knowledge domains and integration of methodologies and workflows related to both computational and experimental tools.
MAMBO is expected to advance the application of data-driven technologies to molecular materials, including predictive machine learning frameworks for materials design and discovery and automated platforms.
\end{abstract}


\end{frontmatter}

%%%%%%%%%%% The article body starts:

\section{Introduction}\label{sec1}
The progress in a wide range of fields in science and technology
has greatly benefited from the development of new tailored
functional and advanced materials, addressing specific needs\cite{Materials2030Manifesto}.
Accordingly, advancements in materials development and manufacturing are considered key sectors for innovation and socio-economical assets\cite{KeyPolicy}.
As well as for several strategic fields\cite{2021DataApplications,QIN2012220}, recent developments of data-driven technologies led to significant progress in research and innovation for materials\cite{Himanen2019Data-DrivenPerspectives,Li2019ADomain,Pollice2021}.
Breakthrough results have also been achieved through the application of multiscale modelling and data-science approaches to materials science and engineering\cite{Agrawal2016Perspective:Science}, also enabled by the advancements in high-performance and high-throughput computing (HPC/HTC), machine learning (ML) and artificial intelligence (AI) in general. The new trend of combining simulations with AI and ML-based techniques depicts a promising future for the discipline\cite{LePiane2020PredictingLearning} where hybrid models, i.e. models seamlessly combining physical-based simulations with AI and ML, can support researchers in the translation of basic concepts and ideas into results and innovation value. Namely, ML and AI technologies can replace computing-intensive steps of simulation protocols, bridging gaps across multi-scale simulations,
or can integrate information on materials at different scales to achieve a comprehensive description of complex systems\cite{Bhatia2021MuMMI}.\\
State-of-the-art approaches
for the design and development of novel materials
are often based on a tight integration between computational and experimental tools. Computational techniques are currently able to tackle a manifold of different and complex scenarios\cite{Rosso2017WhatVersion}, overcoming typical limitations of purely experimental tools. Besides, computational tools also give the possibility to investigate physical and chemical phenomena across a broad range of spatial and temporal scales, enabling the understanding of the interplay between materials properties at different levels\cite{Buehler2013-ab}.
From the experimental point of view, research workflows typically employ a variety of methodologies for gathering information on materials across the entire development process.
However, the vast amount of data typically produced by both computational and experimental approaches is often unstructured and uncorrelated, due to the lack of shared operational strategies for the organization of knowledge and information on materials.
In addition, the volume of data related to materials design and engineering is growing at a fast pace, leading to a strong need to organize and structure information and knowledge\cite{Mavračić2021}.
The efficient structuring of relevant data is at the basis of the progress for accelerating research on materials through automated and predictive platforms and AI.
In this context, initiatives related to the FAIR (Findable, Accessible, Interoperable, Reusable) requirements are currently been pursued to push the development of materials towards a more integrated and structured use of data\cite{Wilkinson2016Comment:Stewardship}\cite{C9SC03766G}\cite{ZHANG20158}.

This strong need for organization of knowledge and information can be
tackled by using ontologies, which have already shown their great potential in the field\cite{Ashino2010MaterialsKnowledge,Cheung2008TowardsMaterials, VOIGT2021129836}.
The term \textit{ontology} is used to define a logical structure that represent concepts, categories, entities, properties and relations of a specific knowledge domain using formal definitions and naming conventions.
Many academic disciplines uses ontology to deal with complexity and to organize raw data into usable information and knowledge. Ranging from the definition of a controlled common jargon to the description of concepts, taxonomies and relations, ontologies have a long history in supporting researchers to organize knowledge in specific domains, enabling faster communication and better design and management of experiments and activities, including complex workflows, processes and procedures. \\
Ontologies are one of the building blocks of semantic technologies
supporting the formal encoding of semantic meanings alongside with data\cite{GRUBER1993199}.
Semantic technologies enable knowledge sharing and discovering, helping to deal with heterogeneous data sources, to develop interoperable platforms, and to extract new information and knowledge from data, such as logical reasoning, which complements statistical learning\cite{Bayerlein2022Perspective, Shuai2017Heterogeneous}.
Moreover, by formally stating the inherent structure of knowledge within a specific subject area, encoding its properties and their mutual relations, new knowledge can potentially emerge, improving current processes and data collection procedures.
Semantic platforms, backed by the cooperation among teams of researchers through ontologies and semantic technologies, can therefore potentially unify research and development efforts in the field of advanced materials and align existing research lines\cite{Horsch2020EMMD}.
In addition, ontologies can provide platforms to
enable interoperability across a broad range of technologies for materials development, integrating investigation and analysis methodologies at different levels of detail and at different scales within unified workflows\cite{Wilkinson2016Comment:Stewardship, Horsch2020ex2019}.

Recent work has been focused on the
application of semantic technologies and methodologies to the development of materials and related applications\cite{Ghedini2017EMMOONTOLOGY, Li2020AnDomain, Hakimi2020, Degtyarenko2008ChEBI:Interest, Horsch2020OntologiesMarketplace, Draxl2018NOMAD:Science, 2021OntologyAssistedModelling}, as in the cooperative effort leading to the Elementary Multiperspective Material Ontology (EMMO)\cite{Ghedini2017EMMOONTOLOGY}.
Stemming from this seminal effort, many domain ontologies tailored for specific use cases have been developed\cite{Li2020AnDomain, Li2019ADomain, Degtyarenko2008ChEBI:Interest}.
Despite these notable endeavors, however, rational strategies for the organization of information and knowledge in several specific sub-fields of materials science and engineering are still missing.
Indeed, previous research in the field has mostly been focused on either generic aspects related to materials and processing (for example, for bulk materials engineering and manufacturing) or materials properties defined at the atomistic level (individual molecules, ideal/periodic crystal structures, etc.)\cite{Ghedini2017EMMOONTOLOGY, Li2020AnDomain, Hakimi2020, Degtyarenko2008ChEBI:Interest, Horsch2020OntologiesMarketplace, Draxl2018NOMAD:Science, 2021OntologyAssistedModelling}.
In this latter case, ontologies are mainly applied to organize information about structure/property relationships obtained from computational investigations at the electronic/atomistic level on materials models.
Other recent work has targeted materials properties at the nanoscale, focusing mostly on the nanosafety and nanotoxicity domains\cite{THOMAS201159}.
Essentially, current available domain ontologies in materials science lack specific capabilities for modelling knowledge from the atomistic/molecular to the nano/micro scale, establishing a link up to the bulk/continuum scale with the support of top/middle-level ontologies such as EMMO.
This is particularly the case of molecular materials, a specific class of functional systems whose basic units are constituted of molecules. The physical and functional properties of molecular materials generally depend on both the structure of their individual molecular sub-units and on the molecular aggregation morphology in condensed phases.
Accordingly, changes in the structure of individual molecules affect the properties of materials at larger (from nanoaggregation to the bulk phase) scales.
In addition, the relationship between molecular structure and aggregation and morphology at different scales is ruled by fabrication and processing conditions, with a dramatic impact on the overall properties of materials\cite{Diao2014MorphologyControl, Fitzner2012CorrelationOligomer, Jones2016SubstrateThinFilm, Virkar2010GrowthMorphology, Kumar2014RelationMorphologyStructure, Lorenzoni2019Graphene}.
Therefore, the comprehensive understanding of the interplay between the structure of molecular units and sub-units,
processing, and aggregation morphology plays a crucial role in the development of molecular materials for
strategic application fields, including organic electronics and optoelectronics (OLEDs, organic thin-film transistors)\cite{Jones2016SubstrateThinFilm, Virkar2010GrowthMorphology, Kumar2014RelationMorphologyStructure, C4RA10954F},
organic and hybrid photovoltaics (organic and hybrid solar cells),\cite{Fitzner2012CorrelationOligomer, Zhang2021ThickFilm, Seri2021Perovskite}
and bioelectronics (neural and brain interfaces, devices based on biomaterials)\cite{Tropp2021DesignBioelectronics, Pradhan2020NatureDerived, Ohayon2020OrganicBioelectronics, Wang2020Bioelectronics2D, Borrachero2019BrainCells}.
The development of applications based on advanced functional materials, and in particular molecular
materials, could therefore greatly benefit from efforts aimed at a more articulated structuring of knowledge in materials science.

The Materials and Molecules Basic Ontology (MAMBO) aims at filling this gap.  MAMBO focuses on a specific domain related to materials science, including molecular materials, nanomaterials, supramolecular materials, molecular thin-films and other related systems. Also, MAMBO targets the improvement of the efficiency of data storage and retrieval infrastructures, merging information obtained from different, heterogeneous sources (for example, integrating computational and experimental tools) with seamless integration. As such, MAMBO is designed to provide an access point for a smoother integration between data-driven technologies and current computational and experimental workflows used to generate materials data, including ML-based techniques.
The approach proposed aims at supporting
the design and development of novel functional materials, which would strongly benefit from a unification of knowledge and information in the field.

The paper is structured as follows:
in Section 2 the relationship between MAMBO and previous related work is discussed;
Section 3 introduces MAMBO and its rationale;
Section 4 presents the details of the development process and applied methodologies;
in Section 5 the implementation of MAMBO is described, as well as the initial tests carried out to ensure the stability and expressiveness of MAMBO, while Section 6 shows how MAMBO can be exploited for research and development in materials science upon a case study in the context of simple workflows, providing an evaluation of the proposed approach; in Section 7 an outline of future perspectives and work is given;
finally, Section 8 draws the conclusions.

\section{Relations and integration with previous work}\label{sec2}
The development of ontologies for the organization of knowledge in fields related to materials design and applications has been the target of recent research efforts\cite{ZHANG20158}.
These efforts resulted in the development of both top/middle level ontologies, to create a general foundation for knowledge organization, and domain ontologies. These latter find application in the organization of knowledge related to specific workflows and procedures related to the development of advanced materials.
In both cases, ontologies have proved to constitute a fundamental base for the future development of interoperable platforms and frameworks for materials.
Most of the work in this context has been collected by the MatPortal initiative, which currently features 21 specialized ontologies\cite{MatPortal}.
The work leading to the development of MAMBO has largely been based on these efforts. Here, we briefly discuss the characteristics of existing ontologies for materials that are more
related to our work.
    \subsection*{EMMO}
    EMMO (the Elementary Multiperspective Material Ontology)\cite{Ghedini2017EMMOONTOLOGY} is the result of a multidisciplinary effort, aimed at the development of a standard representational ontology framework based on current materials modelling and characterization knowledge. Rather than starting from general upper level abstract concepts, as done by other ontologies, the development of EMMO has been based from the very bottom level of practical concepts, using the actual picture of the physical world coming from applied sciences, and in particular from physics and materials science.
    While being extremely useful as a starting point and as a common root, EMMO is (by design) general and non-specific, and can be used as a top- and middle-level ontology.
    The specific role of EMMO is to focus and organize the knowledge of fundamental physics, chemistry and ontology (in the philosophical meaning) behind materials science and, more generally, applied science.
    The purpose of EMMO is to define all the basic building blocks that are necessary for organizing knowledge in the aforementioned scientific domains, while leaving to subsequent work the burden to actually define functional and enforceable categories and classes for practical applications.
    From this common parent, many different sub-ontologies have spawned, some of which we will discuss in the following.
    In the development of MAMBO, we mutuated from EMMO general design criteria and approaches, especially in the definition of relationships between classes, with the purpose of a possible use of MAMBO as a domain ontology of EMMO.
    \subsection*{MDO}
    MDO (Materials Design Ontology)\cite{Li2020AnDomain} defines concepts and relations to cover knowledge in the field of materials design and aims to define a knowledge representation that can unify experimental and computational results within a common framework. MDO is specifically designed for empowering information retrieval from databases collecting information in the computational materials science domain.
    Although potentially able to represent in detail knowledge in several sub-fields of the materials design domain, and especially computational tasks, MDO is principally focused on crystalline/periodic systems and single molecules. As such, MDO is less suited to represent and manage information about non-crystalline compounds and materials (e.g. molecular materials).
    MDO is structured upon Competency Questions\cite{Ontology101},
    answered by specialists in materials science, and Use Cases\cite{Ontology101},
    which define the aim and the foundations on which the ontology is actually designed. MDO is structured as a modular ontology, built around the core concepts of Structure, Provenance and Property.
    The term {\em Structure} defines the basic chemical structure of a specific material,
    while the term {\em Property} defines the actual chemico-physical properties of a material (both quantitative and qualitative).
    Namely, we reused these general concepts defined in MDO, reimplementing them in the context of the extension towards the target domain. As we will show later, we also linked in MAMBO some specific classes and attributes already defined in MDO.
    \\
    \subsection*{DEB}
    DEB (the Device, Experimental scaffolds and medical Device ontology)\cite{Hakimi2020} is an open resource for organizing information about biomaterials, their design, manufacture, and biological testing. It is developed using text analysis for identifying ontology terms from literature on biomaterials,
    systematically curated to represent the domain lexicon.
    DEB may be used for searching terms, performing annotations for machine learning applications, standardized meta-data indexing, and other cross-disciplinary data exploitation.
    Being an ontology for Biological materials, Devices and Experimental scaffolds, DEB is mainly linked to MAMBO from the point of view of foundations and approach.
    Moreover, the complexity and the very heterogeneous nature of the literature and operational approaches are very similar to those addressed by MAMBO.
    One of the interesting peculiarities of DEB consists in the semi-automatic selection of field names to be inserted in the ontology, as opposed to common approaches relying on domain experts. Moreover, the DEB terms set is dynamic, as new elements are added when the need for new terms emerges from new pieces of research. Finally, DEB is overtly oriented towards the integration of machine learning techniques with classical computational approaches. Namely, terms included in DEB have been selected via a bag-of-words approach, and the completeness of the selected terms set has been validated in two different phases: first, a curated selection of articles have been mined, verifying if all articles were easily findable with the terms present in the ontology; then, 70 domain experts have been asked to do the same with their articles and themes of interest, proposing new missing terms to be added to the ontology. The approach pursued in the development of DEB points therefore to self-updating ontologies, which can automatically be upgraded with newer terms and vocabularies leveraging some automation and data-mining techniques.

    \subsection*{ChEBI}
    ChEBI (Chemical Entities of Biological Interest) \cite{Degtyarenko2008ChEBI:Interest} is an ontology, paired to a database, for molecular entities focused on ''small'' chemical compounds.
    ChEBI defines the term ''molecular entity'' as any "constitutionally or isotopically distinct atom, molecule, ion, ion pair, radical, radical ion, complex, conformer, etc., identifiable as a separately distinguishable entity".
    The molecular entities considered are either natural or synthetic potentially bioactive products.
    The concepts defined in ChEBI for the definition of the structural features of molecular systems are therefore particularly useful for the organization of knowledge in domains where individual molecules and their components are involved.
    We took inspiration from ChEBI's approach to the definition of molecular systems and sub-units to develop concepts related to individual molecules, particles and all the respective properties and hierarchies.

    \subsection*{CHMO}
    CHMO, the CHemical Methods Ontology\cite{CHMO}, describes methods used to collect data in chemical experiments, including characterization (mass spectrometry, electron microscopy), materials processing and isolation (ionisation, chromatography, electrophoresis) and to synthesise and fabricate materials (epitaxy,  continuous vapour deposition).
        It also describes the instruments used in these experiments, such as mass spectrometers and chromatography columns.
        It is intended to be complementary to the Ontology for Biomedical Investigations (OBI)\cite{Bandrowski2016OBI}.
    CHMO is a very mature ontology for experimental procedures.
    We reused sections of CHMO for complementing the general structure of MAMBO with classes and relationships related to experimental materials development.

    \subsection*{MSEO}
    Another important ontology for materials science is MSEO: the Materials Science and Engineering Ontology\cite{MSEO}. It uses the Common Core Ontologies stack\cite{Otte2019} and strive to provide material scientists the ability to represent their experiments and related data. MSEO aims at supporting researchers with semantic data management tools, with both human- and machine-readable capabilities, and that can be easily integrated into other scientific domains.

    \subsection*{Other related ontologies and projects}
    MAMBO also relates to other existing projects, platforms, material databases:
    \begin{itemize}

        \item[$\bullet$] The Virtual Materials Marketplace Project (VIMMP)\cite{Horsch2020OntologiesMarketplace} provides an easily accessible, user-friendly hub to access all tangible and intangible components, such as information, knowledge, services and tools to support the efficient decision making, uptake and effective use of materials modelling by a wide range of manufacturing end-users, thereby facilitating an accelerated speed of development and market deployment of new materials. VIMMP proved to be an useful resource for re-using concepts, structures and relations.

        \item[$\bullet$] The Open Databases Integration for Materials Design (OPTIMADE)\cite{Andersen2021OPTIMADESpecification} is a consortium which aims to make materials databases interoperable by developing tools as for example a specification for a common REST API\cite{Andersen2021OPTIMADEAPI}.

         \item[$\bullet$] NOMAD (NOvel MAterials Discovery) \cite{Draxl2018NOMAD:Science,2021OntologyAssistedModelling} is a database that creates, collects, stores, and cleanses computational materials science data, computed by the most important materials-science codes available today. Furthermore, the projects develops tools for mining these data for finding structure, correlations, and novel information that could not be discovered by analysing smaller data sets.

    \end{itemize}

\section{The MAMBO approach and rationale}\label{sec3}
The MAMBO approach stems from a manifold of practical needs in domains related to the development and application of advanced materials, and in particular of molecular materials.
One of the most relevant issues in the field concerns the need for a structured and standardised way to design and implement R\&D activities, considering both materials data, information and workflows.
Targeting practical applications for common research-oriented activities, the MAMBO development approach is focused on the definition of a specific sets of concepts, relations, and tools, to support R\&D on materials. Indeed, materials science requires intrinsically multidisciplinary efforts involving researches with different backgrounds, including physics, chemistry, engineering, computer science.
In this multidisciplinary context, terminologies for the definition of research objects and procedures are unevenly scattered.
However, additional concerns other then terminological ones may arise. For example, the definition of enabling conditions for a given experiment, in terms of required initial input data and conformed outputs, may be needed.
Furthermore, research activities may require several interrelated steps to be completed. These steps can be modelled decomposing complex tasks into subtasks.

To deal with aspects related to the automation and optimization of processes and procedures, we decided to apply concepts mutuated from the Problem-Solving Method (PSM) approach throughout the development of MAMBO.
PSMs are useful tools to implement task-based frameworks in Knowledge Engineering (KE) and can therefore link the representation of knowledge and information to operational tasks\cite{STUDER1998161}.
Accordingly, our approach was guided by the realization of a framework where tools are modelled as PSMs exploiting a domain ontology, which provides  terms and concepts for the specification of tasks, data and workflows\cite{GRUBER1993199, Mizoguchi1995OntologyReuse}.\\
Essentially, the the three basic ingredients of PSMs are the following:
\begin{itemize}
    \item Competence: the description of the input and output behavior related to a given task, together with a description of what the PSM can achieve.
    \item Operational specification: the description of the reasoning process, which links the required knowledge to the specified competence.
    \item Requirements/assumptions: the description of the domain knowledge needed by the PSM to achieve the competence. Simply put, \textit{requirements} and \textit{assumptions} describe the pre-requisites needed by the inference steps described by the operational specification for the application of the PSM to achieve a target.
\end{itemize}
Therefore, PSMs can be used to define and accomplish tasks by applying domain knowledge. This means that the external context of a PSM is made of two pieces: the task to be realized and the domain knowledge that is needed.
The execution of tasks can be defined in terms of methods, which are also related to pre- and post-conditions. Complex operational workflows can be defined in PSMs as a pre-constituted or an incremental (\textit{on-the-fly}) composition of methods\cite{GRUBER1993199, Mizoguchi1995OntologyReuse}.
%In this work, we applied these general principles behind PSM approaches to the development of MAMBO.
This implies that the definition of an ontology can support the organization of knowledge required to apply PSM methods to a specific domain and, at the same time, the definition of PSMs requirements can be used as the starting point for shaping an ontology related to the domain of knowledge.
Following this approach, MAMBO targets a "lightweight ontology"\cite{Benjamins2000KnowledgeSystemTO}, specifically oriented to solve practical use-cases and to organize the required knowledge, rather than developing an all-inclusive ontology for materials science and/or molecular materials.
This approach gives us the opportunity to focus on the real use-cases emerging from research activities of the considered domain.
From the practical point of view, we can use the categories and relations defined in MAMBO to shape methods connected to individual tasks, into which complex problems are decomposed. In turn, we can further integrate in MAMBO concepts and relationships emerging from the analysis of new applications and problems, thus tailoring the ontology towards the solution of real-life tasks.\\
The development of MAMBO requires therefore the definition of typical
tasks (goals and methods) and sub-tasks, methods and pre/post-conditions related to use-case scenarios within the domain considered.
The instances of the pre-conditions will be related to ontology terms. Furthermore, relationships between concepts will be formalized for specific cases.
A thorough analysis of the application scenarios and use-cases is therefore a fundamental prerequisite for the development of MAMBO.\\

    \subsection{Aims and applications}
    MAMBO targets typical situations emerging in the development of molecular materials and akin systems. The design of MAMBO was therefore guided from the analysis of use cases in the context of experimental and computational workflows, as it will be shown below. The generic development workflows and procedures considered can imply very complex tasks, involving knowledge, processes, information and data at different levels. This situation is particularly relevant in multi-scale experimental and computational investigations\cite{Kovachki2022MultiscaleModeling, Fish2021MesoscopicMultiscale, Barreiro2019MultiscaleSilk}, which are of crucial relevance for the development of molecular materials\cite{Mondal2021}.
    An ontology for representing knowledge within this domain should therefore be able to cover aspects related to the properties of materials at different spatial scales and to the properties of processes at different degrees of definition and granularity.
    In this context, workflows define and implements the steps that lead from a specific problem to the solution in terms of computational and/or experimental tasks.
    For example, MAMBO aims at representing the knowledge required, in terms of processes and data structures, to compute the properties of molecular aggregates starting from information on individual molecules only. Here, the initial information (for example, the structure of an individual molecule) is linked to the resulting output (for example, structural properties of a molecular aggregate in the condensed phase) through a complex sequence of individual tasks (i.e. retrieving the structure of a single molecule file from a data source, finding the optimal molecular structure upon a given interaction potential, building a model of a molecular aggregate according to a given simulation protocol, performing a simulation to model the morphology of the aggregate, etc.).
    Ontologies can supporting the design and implementation of complex workflows, assisting the organization of knowledge and the structuring of relevant data. This step enables the realization of high-throughput and automation strategies for the generation of structured materials data, compliance with the FAIR requirements
    and the implementation of efficient data-driven models.
    These latter include AI and ML methods for property screening and prediction, design of materials and process optimization across a broad range of scales. In addition, the semantic interoperability provided by the MAMBO approach will give researchers the possibility to implement platforms for integrating data and knowledge originating from both simulations and experiments.\\

\section{Ontology development}\label{sec4}
    \subsection{Preliminary steps and use cases}
    First of all, we analysed the current literature in the field of ontology development in the materials science and development domain. This analysis led to the links with the ontologies mentioned above,
    while also highlighting the requirements of a specialized domain ontology in the field targeted by MAMBO.
    After this analysis step, we initially adopted a bottom-up approach in the development of the ontology, implementing strategies to identify possible real use-case scenarios where the use of MAMBO can support the organisation of knowledge.
    To this end, we identified a work group of about 10 domain experts, with competences in computational and experimental aspects of materials research and development, supported by knowledge engineers.
    We first asked the experts to describe typical operations, processes, objectives and
    goals related to their research activities, daily routine and tools used, gathering
    a set of relevant tasks and objectives.
    The answers provided by the group of experts can be qualitatively summarized in terms of the following use cases:
    \begin{enumerate}
        \item Retrieval and generation of structured information about molecular materials and related systems.
        \item Definition and implementation of new, complex workflows for the development of molecular materials and related systems.
    \end{enumerate}
Accordingly, one of the main objectives of MAMBO consists in supporting interoperability in the implementation of workflows within the specific domain considered and for the organization of structured data.
MAMBO should thus be able to support the following use case scenarios:

    \begin{itemize}
        \item[$\bullet$] Representing knowledge on integrated modelling and characterization workflows for advanced materials, processes and related technologies.
        \item[$\bullet$] Providing a standard representation of materials modelling and characterization workflows.
        \item[$\bullet$] Providing interoperability across modelling and characterization tools (for example, modelling software tools, characterization workflows, etc.).
        \item[$\bullet$] Providing a basis for the development of tools with search/query capabilities in the field of materials modelling and characterization.
        \item[$\bullet$] Connecting with existing knowledge in the materials science domain.
    \end{itemize}

    \subsection{Competency questions}
    In agreement with other common ontology development schemes\cite{Ontology101}, the initial MAMBO draft
    was built also on the basis of competency questions (CQs)\cite{Ontology101}.
    CQs are usually expressed in natural language and aim at assisting ontology engineers in
    defining the scope of the knowledge considered\cite{Ontology101}.
    Essentially, the ontology should contain all the relevant knowledge needed to
    answer the CQs considered.
    Supported by the domain expert team, we identified an initial set of typical questions for which the information and organization in MAMBO should provide answers, including the following:
    \begin{itemize}
        \item[$\bullet$] What is the chemical formula of the molecules constituting the basic units of the material considered?
        \item[$\bullet$] Is the material made of a single molecular component or is it a blend?
        \item[$\bullet$] Is the material homogeneous at the molecular scale?
        \item[$\bullet$] What are the possible compositions of materials with computed properties (e.g. computed density) falling within a given range?
        \item[$\bullet$] What are the material properties and their values that are produced by a given materials simulation?
        \item[$\bullet$] What are the molecular structures of the interface between two given materials for which a computed property (e.g. computed density) falls within a given range?
        \item[$\bullet$] What is the method used for calculating a given material property?
        \item[$\bullet$] Which are the parameters used in a materials simulation and their respective value ranges?
        \item[$\bullet$] What is the method used for evaluating experimentally a given material property?
        \item[$\bullet$] For a calculated material property, which software produced the simulation results?
        \item[$\bullet$] What are the input and output structures of a materials simulation?
        \item[$\bullet$] What are the input and output structures of a materials characterization?
        \item[$\bullet$] Who are the authors of the simulation for a computed property?
        \item[$\bullet$] Who are the authors of the measurement for an experimentally-measured property?
    \end{itemize}

    Together with the UCs, from CQs we also gathered a first list of relevant words, which has been iteratively updated during the whole process.

    \subsection{Sketching MAMBO}
    In this phase we embraced an “hybrid” approach (bottom-up and top-down) to better represent the different nature of concepts involved in the development of the MAMBO ontology. First, we built an initial set of qualitative relationships among the emerged terms. Then, we defined the ontology classes, also refining their mutual relationships.
    As stated above, we pursued an essentially modular approach, based on a set of
    core concepts and relationships, and subsequently considered their extension to further specialize the structuring of the knowledge within the domain.
        \subsubsection{MAMBO main concepts and core structure} \label{main_concepts}
\begin{figure}[!ht]
        	\centering
        	\includegraphics[scale=1.0]{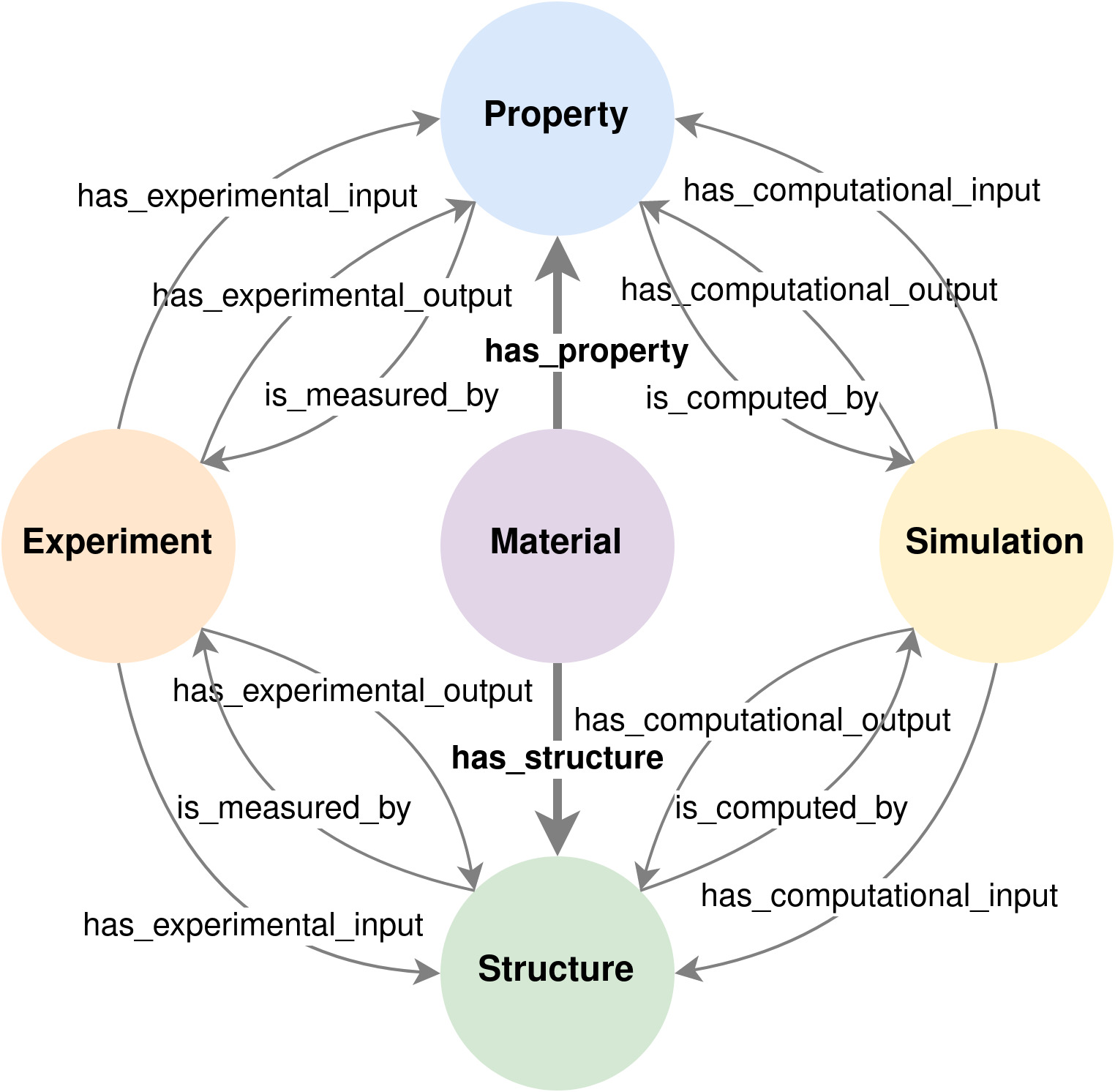}
        	\caption{MAMBO main core classes and relationships: the ontology revolves around the concepts of {\tt Material}, {\tt Simulation} and {\tt Experiment}. An object ({\tt Material}) is represented by its structural features ({\tt Structure}) and properties ({\tt Property}), while computational ({\tt Simulation}) and experimental ({\tt Experiment}) workflows are connected through a common interface to {\tt Property} and to {\tt Structure}.
        	}
        	\label{core}
            \end{figure}

        The entities constituting the core of MAMBO are strongly related to the tasks and use cases cited above: the main concept in MAMBO is that of {\bf material}.
        This is linked to the concepts of {\bf structure} and {\bf property}, which are in turn linked to the concepts of {\bf experiment} and {\bf simulation}.
        This structuring reflects the main recurring terms emerging from the use cases targeted by MAMBO within the specific domain considered.
        Each of these qualitative concepts is strongly connected to at least one of the tasks related to the domain targeted by MAMBO.
        As we will show below, the actual implementation of MAMBO will strongly be based upon the concepts and
        relationships defined at this stage.
        A more detailed definition of this first set of concepts is provided below:
        \begin{itemize}
            \item[$\bullet$] {\bf Material}: the most general concept related to a material, which defines a portion of matter with some specific attributes (kind, quality, etc.)\cite{Ghedini2017EMMOONTOLOGY}.
            Despite very generic in principle, the range of the specific materials covered by MAMBO is narrowed by the set of attributes considered.
            The {\bf Material} concept is related to the following two concepts: {\bf Structure} and {\bf Property}.
            \item[$\bullet$] {\bf Structure}: term representing all the concepts and relations connected to the structural characteristics of a {\bf Material}.
            Sub-hierarchies of the {\bf Structure} concept will specify knowledge related to typical material components within the targeted domain (atoms, molecules, etc.).
            \item[$\bullet$] {\bf Property}: concept representing all the chemical and physical properties of a material (mechanical, electro-magnetic, etc.).
            The {\bf Property} hierarchy is conceptually straightforward: new useful categories representing properties emerging from related and/or neighboring domains can easily be added to the ontology.
            \item[$\bullet$] {\bf Simulation}: concept representing all the information related to a computational workflow leading to a specific result, from the scientific motivations down to the actual software and parameters. As it will be shown below, the hierarchies connected to the {\bf Simulation} concept are, together with {\bf Experiment}, the most complex of the entire ontology, as they should be able to represent a broad range of scenarios related to different simulation tools and protocols.
            \item[$\bullet$] {\bf Experiment}: analogously to {\bf Simulation}, this term represents all the processes and procedures needed to perform an empirical experiment. Together with {\bf Simulation}, the {\bf Experiment} concept represents the knowledge required to define the procedures providing the value of a {\bf Property}, the characteristics of a {\bf Structure} and/or their mutual link. Consequently, {\bf Experiment} and {\bf Simulation} can be viewed as different categories of methods providing different representations of given physical phenomena.
        \end{itemize}
        These concepts are deeply interconnected, as a consequence of the strong mutual relationship among them occurring within typical activities related to materials
        investigation.
        Accordingly, the core structure of MAMBO was designed by associating the main terms of interest described above to ontology classes ({\tt Material}, {\tt Simulation}, {\tt Experiment}, {\tt Structure} and {\tt Property}, respectively). The link between classes was defined through relationships, which also emerged from the analysis of computational and experimental workflows addressing structure and property of materials. A sketch of the basic MAMBO core classes and relationships is depicted in Fig. \ref{core}.

    \subsection{Follow-up questionnaires}
    After the first sketching of MAMBO main structure and relationships, we ran through another round of interviews with domain experts
    in the materials development field.
    We asked computational and experimental domain experts to fill a questionnaire aimed at validating the effectiveness of MAMBO in representing typical tasks and workflows related to the materials development process.
    The interview was based on a set of very generic questions about
    protocols, requirements and use of information and data for
    research on advanced materials.
    This process helped us assessing the main technical requirements of
    experimental and computational workflows in the molecular
    materials domain, thus evaluating the general applicability of
    the core structure of MAMBO in real-life scenarios.
    It is worth noting that in this step we checked the consistency between the MAMBO core concepts and the materials development process
    within the domain considered. Moreover, the answers to the
    questionnaires provide valuable hints to sketch the deeper
    hierarchies of MAMBO, as it will be shown below.
    The questionnaire has been designed in order to offer experts the possibility to answer at different levels of detail.
    The questions posed to experts are the following\footnote{Closed questions have the possible answers reported inside parentheses}:
    \begin{itemize}
        \item[$\bullet$] Briefly describe your fields of study and interest.
        \item[$\bullet$] Which are the fundamental components of your work? (molecules, molecular aggregates)
        \item[$\bullet$] How do you usually structure your research process?
        \item[$\bullet$] Which of the aforementioned steps is usually the most difficult and time consuming?
        \item[$\bullet$] Which role do data have in your research activities?
        \item[$\bullet$] How big is the volume of data you usually have to manage? (Some megabytes, some gigabytes, many gigabytes, terabytes, other).
        \item[$\bullet$] Are data automatically collected? (yes, no, partially).
        \item[$\bullet$] If data are automatically collected only partially, explain which are the automated and manual steps, respectively, of the data acquisition process.
        \item[$\bullet$] Which is the rate of acquisition of data? How often are they collected? How long is this process?
        \item[$\bullet$] Redact a list of terms which are needed to describe your research activities.
        \item[$\bullet$] If a search engine for materials science existed, which kind of queries would you like to be able to do?
        \item[$\bullet$] Fully explain a particularly relevant experiment, from the design phase to the data collection phase.
    \end{itemize}

    We found that the vast majority of the terms and concepts emerged during this process can be mapped onto the core structure of MAMBO.
    We also found that concepts used for the description of specific computational and experimental workflows and protocols require a deeper level
    of specialization for the ontology.
    The specialized hierarchies of MAMBO were initially implemented for a set
    of specific cases, as those emerged from the interviews with
    experts. To this end, extensions of concepts and relationships
    shown in Fig. \ref{core} were considered.
    However, the modular structure of MAMBO enables in principle
    extension towards a broader set of specialized topics.

    \subsection{Applicability and updates}
    In agreement with the design criteria and use cases outlined above, MAMBO is expected to be used by interdisciplinary teams of computational and experimental materials scientists and data scientists to develop and implement workflows for materials design and development, including ML/DL approaches.
    Moreover, MAMBO is expected to be used by interdisciplinary teams of materials scientists to collect and generate structured materials data
    and/or querying databases.
    Initial updates and maintenance will be carried out by the computational and data science teams involved in the development
    and use of MAMBO, supporting new versions on the basis of the needs related to modelling and experimental improvements and breakthroughs.
    The modularity provided by the approach pursued in the development of MAMBO is expected to activate a continuous integration
    process, expanding the scope of the ontology towards neighboring domains and involving a wider community
    in a bottom-up process.

\section{Implementation}\label{sec5}
The structure of MAMBO discussed above was implemented in OWL\cite{Antoniou2004WebOWL}.
The implementation of MAMBO was assisted by Protégé\cite{Protege2018}, a mature, feature-full
and cross-platform software for the implementation of ontologies.
The current OWL implementation of MAMBO is publicly accessible on GitHub\footnote{https://github.com/daimoners/MAMBO}.

We initially implemented the core concepts and relationships shown in Fig. \ref{core}, gradually expanding MAMBO to cover topics related
to specific case studies.
The definition of MAMBO core classes and relationships follows closely the concepts outlined in section \ref{main_concepts}.
A more formal definition of MAMBO core classes is provided below.
\begin{itemize}
    \item[$\bullet$] {\tt Material}: is the main class of MAMBO
    and is related to the {\tt Property} and {\tt Structure} classes via the {\tt has\_property} and {\tt has\_structure} relationships, respectively. An instance of this class represents the abstraction of a given material, and will be related to potentially many instances of
    {\tt Structure} and {\tt Property}, which in turn can be linked to different instances of the {\tt Simulation} or {\tt Experiment} classes.

    \item[$\bullet$] {\tt Structure}: an instance of {\tt Structure} represents the {\em specific} structural characteristics of a {\em specific} occurrence of a {\tt Material}, being it a real world physical material or the object of a computational workflow.
    The {\tt Structure} class is linked to the {\tt Simulation} class via the {\tt is\_computed\_by} relation and to the {\tt Experiment} class via the {\tt is\_measured\_by} relation.

    \item[$\bullet$] {\tt Property}: this class defines the general
    properties of materials, either computed, measured experimentally or
    both.
    Same as for {\tt Structure}, an instance of the {\tt Property} class represents the specific property related to a specific case.
    Moreover, the {\tt Property} class is linked to {\tt Simulation} and {\tt Experiment} via the relationships {\tt is\_computed\_by} and {\tt is\_measured\_by}, respectively, thus in analogy with the
    {\tt Structure} class.

    \item[$\bullet$] {\tt Experiment}: this class is linked
    to {\tt Structure} and {\tt Property} through two corresponding relationships, namely {\tt has\_experimental\_input} and {\tt has\_experimental\_output}, respectively.
    These relationships describe the {\tt Structure} or a {\tt Property}
    connected to a specific experiment in terms of input and/or the
    ability to produce a result.

    \item[$\bullet$] {\tt Simulation}:
    this class essentially mirrors the {\tt Experiment} class on the
    side of computational workflows. The {\tt Simulation} class is linked to {\tt Structure} and {\tt Property} via {\tt has\_input} and {\tt has\_output} relations.

\end{itemize}

\subsection{Deeper hierarchies, relationships and attributes}
\begin{figure}[!ht]
    	\centering
    	\includegraphics[scale=0.40]{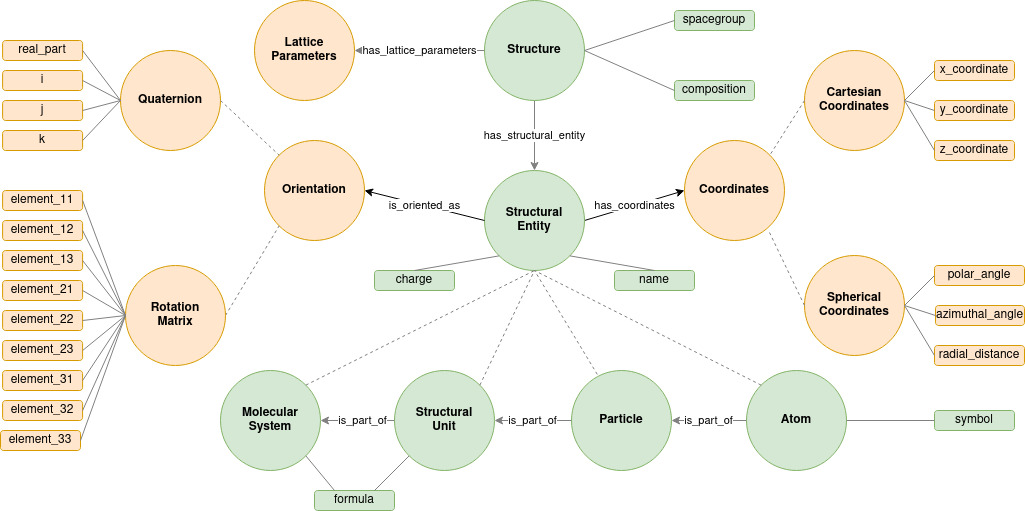}
    	\caption{
    	Draft scheme of the {\tt Structure} class. The main concepts and relationships used in the {\tt Structure} class emerge from the analysis of actual workflows in typical problem solving tasks involving molecular materials. Terms and relationships are connected to both computational and experimental techniques and methods.
    	Circles represent classes; squares represent some examples of attributes for the
    	respective classes; dashed lines represent subclass relationships;
        arrows correspond to defined relations between classes, with the name of the relation written on top.
    	For the sake of clarity, we omitted {\tt Structure} direct subclasses, for example {\tt MolecularAggregate} and {\tt Crystal}.
        Green circles represent the classes directly related to structural entities, i.e. those directly related to {\tt Structure} itself and {\tt StructuralEntity}; orange circles represent the auxiliary classes which encode specific characteristics or spatial features, like all kinds of {\tt Coordinates} types or {\tt Orientation} subclasses, information related to the lattice ({\tt LatticeParameters}) and so on.}
    	\label{structure}
    	\end{figure}

        The core concepts described above and the respective ontology classes as
        sketched in Fig. {\ref{core}} can be further structured into deeper hierarchies and other related concepts. We provide here an overview of the structuring of
        the main MAMBO classes in the initial implementation.
        A comprehensive description of all MAMBO classes and relationships is provided in the MAMBO implementation files.
        As stated above, the {\tt Structure} class aims at defining the structural attributes of materials, with an initial focus on molecular systems and aggregates.
        Accordingly, the class features general attributes pertaining to materials structures (e.g., {\tt spacegroup}, {\tt lattice} and {\tt composition}).
        However, the hierarchy of relations in this class is strongly based on the intrinsic characteristic of materials and chemical entities, which
        are in general composed by many different entities at different scales, ranging from elementary particles and atoms up to molecular aggregates and blends of different materials.
        Following the strongly operative nature of MAMBO,
        the entities representing the constituting units and sub-units
        of materials have been conceptualized on the same level.
        These entities are represented by the {\tt StructuralEntity} class,
        which is linked to the {\tt Structure} class via the {\tt has\_structural\_entity} relation.
        The {\tt StructuralEntity} class is further structured into sub-classes
        (namely {\tt Atom}, {\tt Particle}, {\tt StructuralUnit}, {\tt MolecularSystem}).
        These sub-classes inherit all the parameters of the
        {\tt StructuralEntity} parent class, can share some common
        attributes (e.g., {\tt formula}) and can feature specific
        attributes (e.g., {\tt atomic\_number}, {\tt symbol}).
        Moreover, the {\tt Structure} class can be structured in more detail
        through direct subclasses which specialize the concepts of the
        structure considered, for example the {\tt MolecularAggregate}
        and the {\tt Crystal} classes, respectively.
        These subclasses represent the macro typologies of materials entities that we intend to represent and use. More categories can be easily added to the
        {\tt Structure} class to extend the scope of MAMBO.
        To make a practical example of the hierarchies within the {\tt Structure}
        class, let us consider a periodic model of a molecular aggregate composed of two different molecular species to be used in computational
        simulations. The structure of the system considered (the aggregate) is an
        instance of the {\tt Structure} class. Each individual molecule
        constituting the aggregate is an instance of the {\tt StructuralEntity}
        class. These entities are further defined in terms of the
        {\tt MolecularSystem} subclass.
        In addition, we defined classes that will be used to link MAMBO to other ontologies for materials science. For example, the {\tt Crystal} class
        was inspired by MDO, which, unlike MAMBO, is more specialized on crystalline structures.
        These links empower a more diffuse re-use of terms, relations and patterns
        and future cross-linking of ontologies for materials.
        Moreover, we also decided to inherit other classes from other related
        ontologies, aiming at a further extension of the scope of MAMBO.
        For example, we inherited from MDO the concept and class of {\tt Coordinates}, used to describe the spatial placement of a {\tt StructuralEntity}.
        However, we decided to define the ontology in more detail, using more specialized classes to represent different kinds of coordinates.
        Currently, we implemented for example the {\tt CartesianCoordinates} class (imported from MDO) and the {\tt PolarCoordinates} class.
        This extension emerges from the technical needs of the workflows
        and related PSM tasks considered to draft the ontology. However, their definition as
        sub-classes of the {\tt Coordinates} class imported from MDO should
        enable applicability also in the context of MDO and othe related ontologies.
        A sketch of the {\tt Structure} class is given in figure \ref{structure}.\\

\begin{figure}[!ht]
    	\centering
    	\includegraphics[scale=0.45]{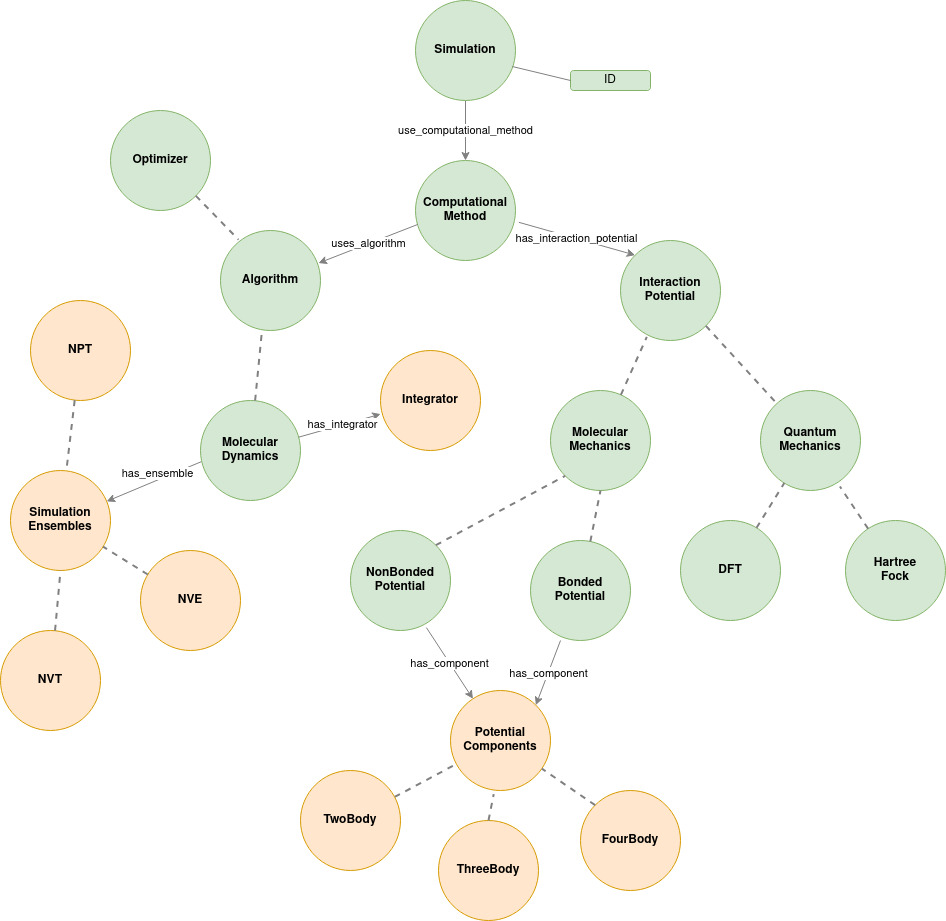}
    	\caption{
    	Scheme of part of the {\tt Simulation} class. The {\tt ComputationalMethod} class gathers the different computational methods and their parameters and is related to the {\tt Algorithm} class (the information about the specific algorithm used for the simulation) and to the {\tt InteractionPotential} class (the information about the interaction potential used in simulations). These classes and their sub-classes are in green. Supplementing classes that collect specific information of algorithms or interaction potentials (e.g., {\tt SimulationEnsemble} and its subclasses) are in orange.
        Arrows corresponds to defined relations between classes, with the name of the relation written on top. For example, {\tt Simulation} and {\tt ComputationalMethods} are connected by a relationship called {\tt use\_computational\_method}. Dashed connections corresponds to the subclass relationships; for example, {\tt Optimizer} is a subclass of {\tt Algorithm}.
    	}
    	\label{simulation}
    \end{figure}

        The {\tt Simulation} and {\tt Experiment} classes aim at defining
        the more general set of workflows and procedures underlying a computational
        or experimental investigation, respectively, on materials.
        Although methodologies, technologies and quantitative characteristics (for
        example, time scales) are quite often very different from each other
        for these two classes, experimental and simulation workflows share several structures and patterns.
        For example, instances of both classes can be labelled according to a unique identifier to be retrievable (namely, an {\tt ID} attribute).
        Notes and information about results of simulations and experiments
        can be stored in a text form, which can be defined by a
        {\tt log} attribute.
        The specific computational or experimental methods related to a simulation
        or an experiment are defined by the
        {\tt ComputationalMethod} and {\tt ExperimentalMethod} class, respectively,
        which include a collection of specific parameters.
        The modular approach pursued enables a swift definition of sub-methods and related parameters in terms of class/sub-class relationships.
        For example, {\tt MolecularDynamics} is a subclass of the {\tt Algorithm} class and {\tt DFT} is a subclass of {\tt QuantumMechanics}, which in turn is a subclass
        of the {\tt InteractionPotential} class, related to
        {\tt ComputationalMethod}.
        A sketch of the {\tt Simulation} class
        is shown in Fig. \ref{simulation}.
        \\
        An analogous structuring can be applied to the {\tt Experiment}
        class, defining the properties and attributes of specific experimental
        methods and tools.
        The {\tt Experiment} class is associated to the {\tt ExperimentalMethod} class (linked through the {\tt use\_experimental\_method} relation) that represents all the possible considered experimental methodologies.
        The sub-classes of {\tt ExperimentalMethod} specialize the
        knowledge on the experimental methods used in the development
        of materials in more detail.
        In the initial implementation of MAMBO, we decided to import
        the subclasses defining specific experimental methods from the
        CHMO ontology\cite{CHMO}.
        The CHMO ontology already features a broad range of experimental
        methods of interest in the development and characterization
        of advanced materials. Other methods of interest can possibly
        be defined in MAMBO through the {\tt ExperimentalMethod} class.

    The {\tt Property} class has general attributes related to values and units,
    which are used to gather quantitative information about materials properties.
    The {\tt Property} class is structured in terms of sub-classes to
    specialize the knowledge about different kinds of materials properties.
    The hierarchical structuring of the {\tt Property} class in terms
    of sub-classes is straightforward.
    This structuring reflects the representation of different chemico-physical properties of materials at a finer resolution, and can include a broad range of materials properties, as done in previous work\cite{Ghedini2017EMMOONTOLOGY}.
    In the initial implementation of MAMBO, we
    implemented explicitly a basic set of materials properties. However, further
    properties can possibly be included to extend the scope of MAMBO to broader
    application domains.

\section{Exploiting MAMBO for Research and Development of Materials}\label{sec6}
\begin{figure}[!ht]
        	\centering
        	\includegraphics[scale=1.8]{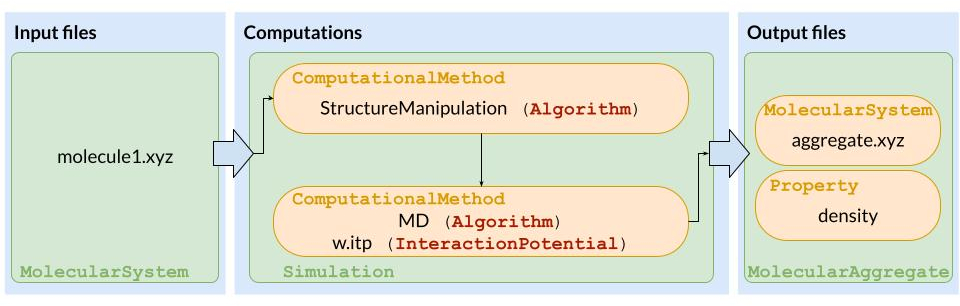}
        	\caption{
         A visual description of the workflow discussed in Sec. \ref{sec6}. The first block contains the input files, which are representable as {\tt MolecularSystem} instances (see also Fig. \ref{mambo_input}) as individuals; the second block consists of all the files and software needed to perform the actual simulation (see Fig. \ref{mambo_comp_method} for more details); finally, the third block represents the output obtained from the simulation, with information about the structure of the molecular aggregate and the resulting computed density.}
        	\label{mambo_workflow_full}
    \end{figure}

    The applicability of MAMBO in the organization of knowledge in the target domain was assessed by analysing simple typical workflows related to R\&D for materials and in particular molecular materials.
    In the following, we focus on simulation workflows for investigations on molecular materials.
    The analysis of simulation workflows, in particular, allows us to define technical requirements and tune accordingly the expressiveness of MAMBO in addressing the specific knowledge involved in the description of materials at different scales (from particles to aggregates).
    Following the PSM approach, a general workflow connecting initial information and conditions (pre-requisites) and final output (post-requisites) is decomposed into tasks and sub-tasks.
    The definition of tasks and sub-tasks and the domain knowledge is organized in terms of the structure provided by MAMBO.
    Let us first consider a simulation workflow for the evaluation of the chemico-physical properties of a molecular aggregate made of identical molecules by molecular dynamics (MD).
    While simple, this workflow exhibits the main features of more complex simulations.
    The consistent representation of this workflow within MAMBO can therefore be instructive of the approach pursued and gives possible hints of the ability to formalize more complex cases.
    This macro-task can be decomposed into several interconnected computational sub-tasks, which involve different operations on structured data.
    From the practical point of view, the overall workflow is generally realized by applying specialized simulation software, which implements specific computational methods, operating on structured input files and producing output files as results.
    Other operations may require the manipulation of files and data structures.
    In the case of the considered workflow, we need for example input files containing information about the structure of the considered molecule.
    This information is further processed by specialized software, implementing computational methods, which provide an output in terms of molecular properties.
    The considered methods can include for example structure manipulation tools (simulation box builders, etc.) and MD specific algorithms for equilibrating molecular aggregates in different conditions\cite{Baldoni2018SpatialOLEDs, Lorenzoni2018NanoscaleMorphology}.
    The workflow produces a structured information containing for example a snapshot of the structure of the simulated aggregate in the considered conditions and/or derived properties (for example, the computed equilibrium density of the aggregate in $kg \ m^{-3}$).
    A sketch of this workflow is shown in Fig. \ref{mambo_workflow_full}.

\begin{figure}[!ht]
        	\centering
        	\includegraphics[scale=2.0]{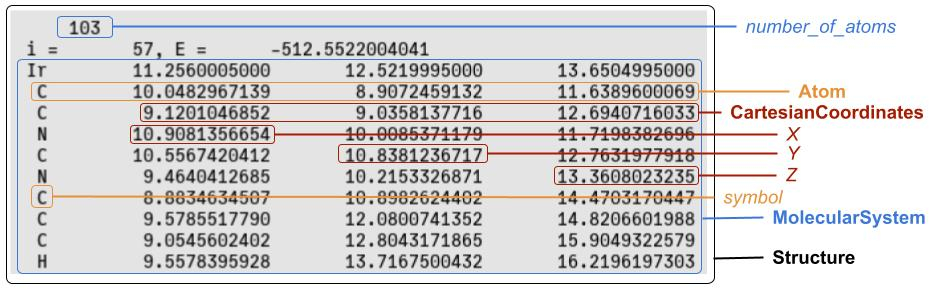}
        	\caption{
         An excerpt of a real-world input file containing structural information about a molecule encoded in the standard xyz format.
        	In particular, the file contains information on the Cartesian coordinates and symbols of all the atoms in the molecule and the total number of atoms. Some of the involved MAMBO instances and class attributes are highlighted in different colors.
        	Black: {\tt Structure} instance; blue: {\tt MolecularSystem} instance; orange: {\tt Atom} instance and attributes; red: {\tt CartesianCoordinates} instance and attributes.
         }
        	\label{mambo_input}
    \end{figure}

\begin{figure}[!ht]
        	\centering
        	\includegraphics[scale=2.5]{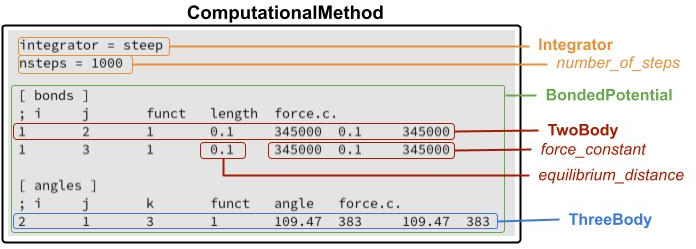}
        	\caption{
         An excerpt of a real-world configuration file containing information about a simulation.
        	This example shows possible encoding in formats used by common software packages for MD simulations (here a syntax mutuated from the Gromacs\cite{BERENDSEN199543} format is considered).
        	In particular, the file contains information about the type of {\tt Integrator}, the definition of the interaction potential used in MD simulations (for example, parameters for bonded potential terms, collected by an instance of {\tt BondedPotential}).
        	Involved MAMBO instances and class attributes are highlighted in different colors.
        	Black: {\tt ComputationalMethod} instance; green: {\tt BondedPotential} instance; blue: {\tt ThreeBody} instance; red: {\tt TwoBody} instance and attributes; yellow: {\tt Integrator} instance and attributes.}
        	\label{mambo_comp_method}
    \end{figure}

    The decomposition of the workflow as sketched in Fig. \ref{mambo_workflow_full} highlights the parallelism between the involved knowledge and instances of MAMBO classes. For example we can identify the following:
    \begin{itemize}
        \item The initial information about the molecular system considered is an instance of the {\tt Structure} class, which is linked to the {\tt Material} class via the {\tt has\_structure} relationship. In particular, the information pertains to the {\tt MolecularSystem} subclass.
        \item More detailed knowledge on the molecular system considered can be structured in terms of instances of the {\tt Atom} class, which contain information about individual atoms of the molecule. In turn, the position of individual atoms corresponds to instances of the {\tt CartesianCoordinates} class.
        \item Information on the tools for the manipulation of data structure and on MD algorithms can be represented as instances of the {\tt ComputationalMethod} class.
        \item In analogy with the input data, part of the information provided by the workflow can be represented as an instance of the
        {\tt Structure} class. In particular, the simulated structure of the molecular aggregate is an instance of the {\tt MolecularAggregate} class.
        \item The computed property of the molecular aggregate (for example, the computed density) is an instance of the {\tt Property} class.
    \end{itemize}
    An example of the parallelism between the structural information on a molecule stored as a file and encoded in a standard format in the context of molecular simulations (xyz format) and corresponding attributes of MAMBO classes is shown in Fig. \ref{mambo_input}. A similar example for attributes of classes pertaining to the {\tt ComputationalMethod} class is shown in Fig. \ref{mambo_comp_method}.


    The link between the structure provided by MAMBO and the data defining a specific computational workflow can be provided by metadata and/or annotations, which can be implemented in a variety of standard formats\cite{TengLV2018JSON}.

    The applicability of MAMBO in the definition of the workflow considered above and defined by exploiting PSM techniques (competences - input/output, operational specifications and requirements) shows the potential of the proposed approach in the context of specific applications in the materials development pipeline. This approach can be easily extended to more complex systems and processes.
    The semantic interoperability ground provided by MAMBO in the materials science domain provides the basic components to represent complex workflows in terms of basic and reusable building blocks enabling high-throughput and automated data processing.

\section{Future work}\label{sec7}
Work is currently in progress to design and realize platforms implementing the use cases targeted initially by MAMBO.
In particular, we aim at applying MAMBO in data-intensive domains related to materials development, both for data retrieval and for storage.
We plan to use MAMBO to organize and formally structure procedures for data management, processing and analysis, avoiding pitfalls both on a performance and on a logical point of view. The automation of data-related pipelines will enable fully automated data management and information retrieval and the univocal and standardized storage of information on materials.
These activities will also be integrated into the more general effort of the materials science community to create common marketplaces and databases for materials development, implementing semantic technologies to define and store information using formally consistent axioms and definitions, thus overcoming current limitations.\\
Moreover, we plan to enforce MAMBO for developing formal methods for defining workflows using PSM-related languages and technologies. Our objective is to contribute to activities targeted to the support of researchers and users by providing a standardized and formal way to describe workflows in the materials research domain\cite{Huber2020AiiDA}.
The interoperable definition of workflows, in turn, enables full automation and reproducibility.
MAMBO classes and relation will be used to define and represent entities related to specific workflows.
The semantic frameworks will be linked to data structures, software, procedures and protocols to define the operational aspects of the workflows and for implementing automated processes.
The link between data and workflows will enable the realization of data-driven frameworks based on semantic technologies, such as decision support platforms and predictive ML tools.
\\
Finally, future activities will focus on the extension of the scope of MAMBO to cover broader aspects of the knowledge involved in the development of materials. This work will be guided by the analysis of more complex workflows involved in specific applications, defined with the support of domain experts. For example, the knowledge related to the development of applications involving molecular materials (devices, sensors, etc.) can be considered.
The increase in the degree of complexity of the workflows considered is expected to lead to the definition of a larger number of subclasses, specializing the knowledge in more detail. In this process, we also expect the possible revision of some components of the overall MAMBO structure, and a tighter interaction and link with other related ontologies.
While very promising in addressing the specific contexts considered, we can already envision some limitations of MAMBO and of the approach proposed, which highlight possible actions targeted to future improvements.
%Future work will be carried out to assess the potential use of MAMBO as a semantic tool to build data-driven platforms for implementing R{\and}D workflows in the materials science domain, specifically tailored to the application of predictive machine learning technologies.

\section{Conclusions}\label{sec8}
In this work, we discussed the main features of MAMBO, an ontology focused on representation and knowledge organisation in the molecular materials domain.
Focusing on operative and data-driven approaches, MAMBO addresses interoperability between the experimental and computational sub-domains, thus potentially empowering data-intensive ML/DL applications.
The structure of MAMBO is based on a broad range of concepts and relationships of common use in the field, including methods and approaches for the multiscale modelling of molecular materials.\\
%We also aim at using MAMBO to implement PSM approaches for real use-case scenarios. In this respect, ontologies provide a valuable semantic platform for defining the entities and objects emerging from the application of PSMs.
%In turn, the application of PSM techniques can provide a feedback for the development of the ontology.\\
The design of MAMBO was guided by the definition of computational and experimental workflows for materials development in the framework of PSM techniques. Here, ontologies provide a valuable semantic platform for defining the entities and objects emerging from the application of PSMs.
The initial development of MAMBO was carried out according to a
%We discussed the process leading to the initial development of MAMBO. The
hybrid (top-down qualitative structuring of concepts and bottom-up formal specialization of knowledge) approach, starting from the identification of the possible use cases, the definition of competency questions and task definition. This phase led to a sketch of concepts, terms and relationships, followed by a formal implementation in OWL.
Instantiation tests demonstrated the effectiveness and expressive power of MAMBO in the case of simple workflows in relation to the targeted domain. Here, MAMBO allowed us to re-construct and define formally the main components of a realistic computational workflow, representing processes, objectives and entities involved through MAMBO classes and relationships. \\

%\begin{acks}
%We thank the group of domain experts who supported us in the development of MAMBO from initial design and sketching to the validation stage and, in particular, researchers at the CNR Institute for Nanostructured Materials in Bologna, Italy.
%\end{acks}

%\section{Declarations}
%Data sharing policies are not applicable to this article as no datasets were generated or analysed during the current study.
%
%\section{Conflict of interest}
%The authors declare that they have no conflict of interest to this work.

%%%%%%%%%%% The bibliography starts:

%%%%%%%%%%%%%%%%%%%%%%%%%%%%%%%%%%%%%%%%%%%%%%%%%%%%%%%%%%%%%
%%                  The Bibliography                       %%
%%                                                         %%
%%  ios1.bst will be used to                               %%
%%  create a .BBL file for submission.                     %%
%%                                                         %%
%%                                                         %%
%%  Note that the displayed Bibliography will not          %%
%%  necessarily be rendered by Latex exactly as specified  %%
%%  in the online Instructions for Authors.                %%
%%                                                         %%
%%%%%%%%%%%%%%%%%%%%%%%%%%%%%%%%%%%%%%%%%%%%%%%%%%%%%%%%%%%%%

\nocite{*}
% if your bibliography is in bibtex format, use those commands:
\bibliographystyle{ios1}           % Style BST file.
\bibliography{bibliography}        % Bibliography file (usually '*.bib')

% or include bibliography directly:
%\begin{thebibliography}{0}
%\bibitem{r1} F. Author, Information about cited object.
%
%\bibitem{r2} S. Author and T. Author, Information about cited object.
%\end{thebibliography}

\end{document}